\documentclass{appolb}
\usepackage{axodraw}

\Preprinttrue

\newcommand{\FA}{{\sl FeynArts}}
\newcommand{\mma}{{\sl Mathematica}}
\newcommand{\FC}{{\sl FormCalc}}
\newcommand{\FO}{{\sl FORM}}
\newcommand{\LT}{{\sl LoopTools}}
\newcommand{\MW}{M_{\rm W}}
\newcommand{\MZ}{M_{\rm Z}}
\newcommand{\lbrac}{\symbol{123}}
\newcommand{\rbrac}{\symbol{125}}

\newcommand{\cpc}[3]{{\sl Comp. Phys. Commun.} {\bf #1} (19#2) #3}
\newcommand{\fp}[3]{{\sl Fortschr. Phys.} {\bf #1} (19#2) #3}
\newcommand{\np}[3]{{\sl Nucl. Phys.} {\bf #1} (19#2)~#3}

\newcommand{\pr}[3]{{\sl Phys. Rev.} {\bf #1} (19#2) #3}
\newcommand{\zp}[3]{{\sl Z. Phys.} {\bf #1} (19#2) #3}
\newcommand{\nim}[3]{{\sl Nucl. Instr. Meth.} {\bf #1} (19#2)~#3}

\begin{document}


\title{Loop Calculations with \\
\FA, \FC, and \LT}

\author{Thomas Hahn\\
\address{Institut f\"ur Theoretische Physik, Universit\"at Karlsruhe\\
D--76128 Karlsruhe, Germany}}

\maketitle

\begin{abstract}
Three programs are presented for automatically generating and calculating
Feynman diagrams: the diagrams are generated with \FA, algebraically
simplified with \FC, and finally evaluated numerically using the \LT\
package. The calculations are performed analytically as far as possible,
with results given in a form well suited for numerical evaluation. The
latter is then straightforward using the implementations of the one-loop
integrals in \LT.
\end{abstract}


\section{Introduction}

With the increasing accuracy of experimental data, one-loop calculations
have long since become indispensible. Doing such calculations by hand is
arduous and error-prone and in some cases simply impossible. So for some
time already, software packages have been developed to automate these
calculations (\eg \cite{MeBD91,xloops}). Yet one remaining obstacle is
that these packages generally tackle only part of the problem, and there
is still considerable work left in making them work together.

In this paper three packages, \FA, \FC, and \LT, are presented which
work hand in hand. The user has to supply only small driver programs whose
main purpose is to specify the necessary input parameters. This makes the
whole system very ``open'' in the sense that the results are returned as
\mma\ expressions which can easily be manipulated, \eg to select or modify
terms.

\FC\ can work either in dimensional regularization or in constrained
differential renormalization \cite{techniques}, the latter of which
is equivalent at the one-loop level to regularization by dimensional
reduction \cite{HaP98}. This makes \FC\ suitable \eg for calculations in
supersymmetric models.

Since one-loop calculations can range anywhere from a handful to several
hundreds of diagrams (particularly so in models with many particles like
the MSSM), speed is an issue, too. \FC, the program which does the
algebraic simplification, therefore uses \FO\ \cite{Ve91} for the
time-consuming parts of the calculation. Owing to \FO's speed, \FC\ can
process, for example, the 1000-odd one-loop diagrams of W--W scattering in
the Standard Model \cite{DeH98} in about 5 minutes on an ordinary Pentium
PC.

The following table summarizes the steps in a one-loop calculation and
the distribution of tasks among the programs \FA, \FC, and \LT\/:
\begin{center}
\begin{tabular}{|c|c|l|l}
\cline{1-1} \cline{3-3}
		&& $\bullet$ Create the topologies \\
Diagram		&& $\bullet$ Insert fields \\
generation	&& $\bullet$ Apply the Feynman rules \\
		&& $\bullet$ Paint the diagrams
& \smash{\raise 4.3ex%
  \hbox{$\left.\vrule width 0pt depth 5ex height 0pt\right\}$ \FA}} \\
\cline{1-1} \cline{3-3}
\multicolumn{1}{c}{$\downarrow$} \\
\cline{1-1} \cline{3-3}
		&& $\bullet$ Contract indices \\
Algebraic	&& $\bullet$ Calculate traces \\
simplification	&& $\bullet$ Reduce tensor integrals \\
		&& $\bullet$ Introduce abbreviations
& \smash{\lower 1.1ex%
  \hbox{$\left.\vrule width 0pt depth 10.5ex height 0pt\right\}$ \FC}} \\
\cline{1-1} \cline{3-3}
\multicolumn{1}{c}{$\downarrow$} \\
\cline{1-1} \cline{3-3}
		&& $\bullet$ Convert \mma\ output \\
Numerical	&& \qquad to Fortran code \\
evaluation	&& $\bullet$ Supply a driver program \\
		&& $\bullet$ Implementation of the integrals
& $\left.\mathstrut\right\}\,$ \LT \\
\cline{1-1} \cline{3-3}
\end{tabular}
\end{center}
The following sections describe the main functions of each program.
Furthermore, the \FC\ package contains two sample calculations in the
electroweak Standard Model, $ZZ\to ZZ$ \cite{DeDH97} and $e^+e^-\to
\bar t\,t$ \cite{BeMH91}, which demonstrate how the programs are used
together.

\section{\FA}

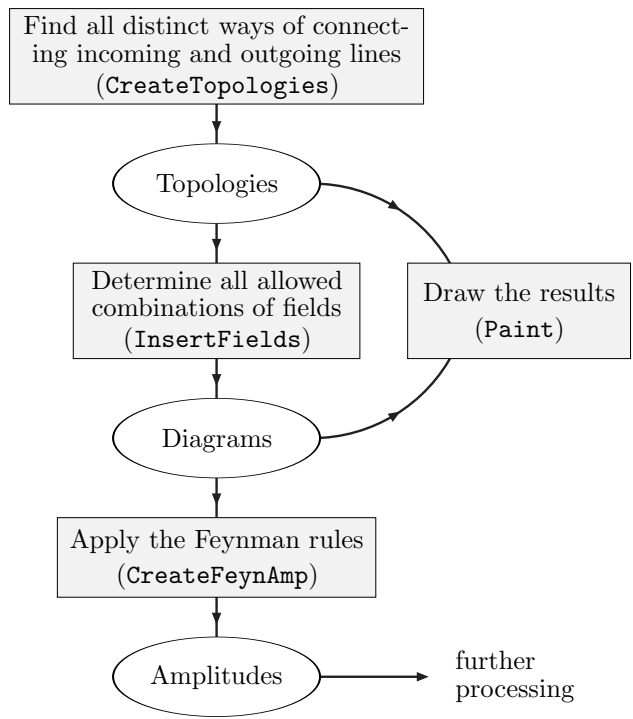
\begin{figure}
\begin{center}
\begin{small}
\unitlength=1bp%
\begin{picture}(230,265)(14,16)
\SetScale{.6}
\SetWidth{1.5}
\ArrowLine(150,411)(150,380)
\ArrowLine(150,335)(150,308)
\ArrowLine(150,251)(150,222)
\ArrowLine(150,175)(150,148)
\ArrowLine(150,101)(150,72)

\Line(200,50)(280,50)
\ArrowLine(279,50)(280,50)
\Text(180,32.8)[lb]{further}
\Text(180,28.8)[lt]{processing}

\ArrowArcn(210,260)(100,90,27)
\ArrowArc(210,300)(100,-90,-27)
\SetWidth{.5}

\GBox(20,410)(280,470){.95}
\Text(90,270.4)[b]{Find all distinct ways of connect-}
\Text(90,260)[b]{ing incoming and outgoing lines}
\Text(90,257.6)[t]{({\tt CreateTopologies})}
\GOval(150,360)(25,65)(0){1}
\Text(90,215.2)[]{Topologies}
\GBox(60,250)(240,310){.95}
\Text(90,174.4)[b]{Determine all allowed\vphantom{p}}
\Text(90,164)[b]{combinations of fields\vphantom{p}}
\Text(90,161.6)[t]{({\tt InsertFields})}
\GBox(270,250)(410,310){.95}
\Text(204,169.6)[b]{Draw the results\vphantom{p}}
\Text(204,165.6)[t]{({\tt Paint})}
\GOval(150,200)(25,65)(0){1}
\Text(90,119.2)[]{Diagrams}
\GBox(50,100)(250,150){.95}
\Text(90,76.8)[b]{Apply the Feynman rules}
\Text(90,72.8)[t]{({\tt CreateFeynAmp})}
\GOval(150,50)(25,65)(0){1}
\Text(90,29.6)[]{Amplitudes}
\end{picture}
\end{small}
\end{center}
\caption{\label{fig:feynarts}%
Flowchart for the generation of Feynman amplitudes with \FA.}
\end{figure}

\FA\ is a \mma\ package for the generation and visualization of Feynman
diagrams and amplitudes \cite{KuBD91}. It works in the three basic steps
sketched in Fig.~\ref{fig:feynarts}.

The first step is to create all different topologies for a given number of
loops and external legs. For example, to create all one-loop topologies
for a $1\to 2$ process, the following call to {\tt CreateTopologies} is
used:
\begin{verbatim}
  top = CreateTopologies[1, 1 -> 2]
\end{verbatim}

In the second step, the actual particles in the model have to be
distributed over the topologies in all allowed ways. E.g.\ the diagrams
for $Z\to b\bar b$ are produced with
\begin{verbatim}
  ins = InsertFields[top, V[2] -> {F[4,{3}], -F[4,{3}]}]
\end{verbatim}
where {\tt F[4,\,\lbrac 3\rbrac]} is the $b$-quark,
{\tt -F[4,\,\lbrac 3\rbrac]} its antiparticle, and \verb=V[2]= the Z
boson. The fields and their couplings are defined in a special file, the
model file, which the user can supply or modify. Model files are currently
provided for QED, the electroweak Standard Model, and QCD; a MSSM model
file is in preparation.

The diagrams can be drawn with \verb=Paint[ins]=, depending on the
options either on screen, or in a PostScript or \LaTeX\ file. Finally,
the analytic expressions for the diagrams are obtained by
\begin{verbatim}
  amp = CreateFeynAmp[ins]
\end{verbatim}

\section{\FC}

The evaluation of the \FA\ output proceeds in two steps:
\begin{enumerate}
\item
The symbolic expressions for the diagrams are simplified algebraically
with \FC\ which returns the results in a form well suited for numerical
evaluation.
\item
The \mma\ expressions then need to be translated into a Fortran program.
(The numerical evaluation could, in principle, be done in \mma\ directly,
but this becomes very slow for large amplitudes.) The translation is done
by the program {\sl NumPrep} which is part of the \FC\ package. For
compiling the generated code one needs a driver program (also in \FC), and
the numerical implementations of the one-loop integrals in \LT.
\end{enumerate}

The structure of \FC\ is simple: it prepares the symbolic expressions of
the diagrams in an input file for \FO, runs \FO, and retrieves the
results. This interaction is transparent to the user. \FC\ combines the
speed of \FO\ with the powerful instruction set of \mma\ and the latter
greatly facilitates further processing of the results. The following
diagram shows schematically how \FC\ interacts with \FO\/: 
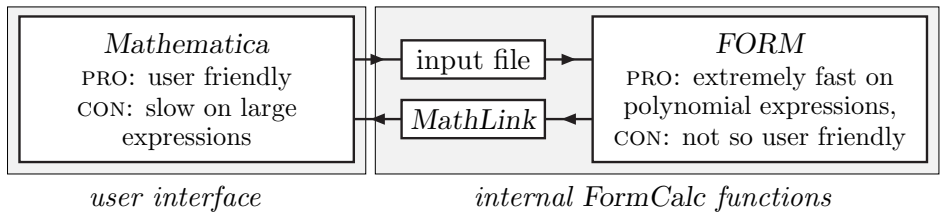
\begin{center}
\unitlength=1bp%
\begin{picture}(351,78.3)(-36,8)
\SetScale{.9}
\SetWidth{0}
\GBox(-40,25)(110,95){.95}
\GBox(114,25)(350,95){.95}
\SetWidth{1}
\BBox(-35,30)(105,91)
\BBox(205,30)(345,91)
\BBox(125,65)(185,81)
\ArrowLine(185,73)(205,73)
\ArrowLine(105,73)(125,73)
\Text(139.5,64.8)[]{input file}
\BBox(125,40)(185,56)
\ArrowLine(205,48)(185,48)
\ArrowLine(125,48)(105,48)
\Text(139.5,43.2)[]{\sl MathLink}
\Text(31.5,72)[]{\mma}
\Text(31.5,58.5)[]{\small {\sc pro:} user friendly}
\Text(31.5,46.8)[]{\small {\sc con:} slow on large}
\Text(31.5,36)[]{\small expressions}
\Text(247.5,72)[]{\FO}
\Text(247.5,58.5)[]{\small {\sc pro:} extremely fast on}
\Text(247.5,47.7)[]{\small polynomial expressions,}
\Text(247.5,36)[]{\small {\sc con:} not so user friendly}
\Text(27,18)[t]{\it user interface}
\Text(207,18)[t]{\it internal \FC\ functions}
\end{picture}
\end{center}

The main function in \FC\ is {\tt OneLoop} (the name is not strictly
correct since it works also with tree graphs). It is used like this:
\begin{verbatim}
  << FormCalc`
  amps = << myamps.m     (* load some amplitudes *)
  result = OneLoop[amps]
\end{verbatim}
where it is assumed that the file {\tt myamps.m} contains amplitudes
generated by \FA. {\tt OneLoop} uses dimensional regularization by
default. To calculate in constrained differential renormalization
($\equiv$ dimensional reduction at the one-loop level), one has to put
{\tt \$Dimension = 4} before invoking {\tt OneLoop}. Note that
{\tt OneLoop} needs no declarations of the kinematics of the underlying
process; it uses the information \FA\ hands down.

Even more comprehensive than {\tt OneLoop}, the function {\tt ProcessFile}
can process entire files. It collects the diagrams into blocks such that
index summations (\eg over fermion generations) can later be carried out
easily, \ie only diagrams which are summed over the same indices are put
in one block. {\tt ProcessFile} is invoked \eg as
\begin{verbatim}
  ProcessFile["vertex.amp", "results/vertex"]
\end{verbatim}
which reads the \FA\ amplitudes from {\tt vertex.amp} and produces
files of the form {\tt results/vertex{\it id}.m}, where {\it id} is an
identifier for a particular block.

{\tt OneLoop} and {\tt ProcessFile} return expressions where spinor
chains, dot products of vectors, and Levi-Civita tensors contracted with
vectors have been collected and abbreviated. A term in such an expression
may look like
\begin{verbatim}
  C0i[cc1, MW2, S, MW2, MZ2, MW2, MW2] *
    ( AbbSum12*(-8*a2*MW2 + 4*a2*MW2*S2 - 2*a2*CW^2*MW2*S2 +
        16*a2*CW^2*S*S2 + 4*a2*C2*MW2*SW^2) +
      Abb47*(-32*a2*CW^2*MW2*S2 + 8*a2*CW^2*S2*T +
        8*a2*CW^2*S2*U) -
      AbbSum13*(-64*a2*CW^2*MW2*S2 + 16*a2*CW^2*S2*T +
        16*a2*CW^2*S2*U) )
\end{verbatim}
Here, the tensor coefficient function $C_1(\MW^2, s, \MW^2, \MZ^2, \MW^2,
\MW^2)$ is multiplied with a linear combination of abbreviations like
{\tt Abb47} or {\tt AbbSum12} with certain coefficients. These
coefficients contain the Mandelstam variables {\tt S}, {\tt T}, and
{\tt U} and some short-hands for parameters of the Standard Model, \eg
${\tt a2} = \alpha^2$.

The abbreviations like {\tt Abb47} or {\tt AbbSum12} are introduced
automatically and can significantly reduce the size of an amplitude. The
definitions of the abbreviations can be retrieved by {\tt Abbreviations[]}
which returns a list of rules such that
{\tt result\,\,//.\,Abbreviations[]} gives the full, unabbreviated
expression.

\section{\LT}

\LT\ supplies the actual numerical implementations of the one-loop
functions needed for programs made from the \FC\ output. It is based on
the reliable package {\sl FF} \cite{vOV90} and provides in addition to the
scalar integrals of {\sl FF} also the tensor coefficients in the
conventions of \cite{De93}. \LT\ offers three interfaces: Fortran, C++,
and \mma, so most programming tastes should be served.

Using \LT\ functions in Fortran and C++ is very similar. In Fortran
it is necessary to include the two files {\tt tools.F} and {\tt tools.h},
the latter one in every function or subroutine. In C++, {\tt ctools.h}
must be included once. Before using any \LT\ function, {\tt bcaini} must
be called and at the end of the calculation {\tt bcaexi} may be called to
obtain a summary of errors. It is of course possible to change parameters
like the scale $\mu$ from dimensional regularization; this is described in
detail in the manual \cite{FCLTGuide}.

\pagebreak[4]

A very simple program would for instance be

\smallskip

\begin{footnotesize}

\setbox77=\vbox{%
\begin{verbatim}
#include "tools.F"

    program simple
#include "tools.h"     
    call bcaini
    print *, B0(1000D0,50D0,80D0)
    call bcaexi
    end
\end{verbatim}}

\setbox78=\vbox{%
\begin{verbatim}
#include "ctools.h"

main()
{
 bcaini();
 cout << B0(1000.,50.,80.) << "\n";
 bcaexi();
}
\end{verbatim}}

\noindent
\begin{picture}(350,100)
\Line(0,0)(0,100)
\Line(0,100)(130,100)
\Line(130,100)(130,90)
\Line(130,90)(170,90)
\Line(170,90)(170,0)
\Line(170,0)(0,0)
\put(150,93){\makebox(0,0)[cb]{Fortran}}
\put(5,100){\makebox(0,0)[lt]{\box77}}
\SetOffset(180,0)
\Line(0,0)(0,100)
\Line(0,100)(148,100)
\Line(148,100)(148,90)
\Line(148,90)(180,90)
\Line(180,90)(180,0)
\Line(180,0)(0,0)
\put(344,93){\makebox(0,0)[cb]{C++}}
\put(185,100){\makebox(0,0)[lt]{\box78}}
\end{picture}
\end{footnotesize}

\smallskip

The \mma\ interface is even simpler to use:
\begin{verbatim}
In[1]:= Install["bca"]
\end{verbatim}
\begin{verbatim}
In[2]:= B0[1000, 50, 80]
\end{verbatim}
\begin{verbatim}
Out[2]= -4.40593 + 2.70414 I
\end{verbatim}

\section{Requirements and Availability}

All three packages require \mma\ 2.2 or above; \FC\ needs in addition \FO,
preferably version 2 or above; \LT\ needs a Fortran compiler,
{\tt gcc}/{\tt g++}, and GNU make.

The packages should compile and run without change on any Unix
platform. They are specifically known to work under DEC Unix, HP-UX,
Linux, Solaris, and AIX. All three packages are open source and stand
under the GNU library general public license. They are available from
\begin{verbatim}
  http://www-itp.physik.uni-karlsruhe.de/feynarts
  http://www-itp.physik.uni-karlsruhe.de/formcalc
  http://www-itp.physik.uni-karlsruhe.de/looptools
\end{verbatim}

\section*{Acknowledgements}

This work has been supported by DFG under contract Ku 502/8--1.

\begin{flushleft}

\end{flushleft}

\end{document}